%% file: main.tex
\documentclass[sigconf]{acmart}

\input{macros}
\input{proj-macros}

\input{eval-macros}
\input{acronyms}

\usepackage{listings}
\usepackage[utf8]{inputenc}
\usepackage[english]{babel}

\usepackage{tablefootnote}

\usepackage{multicol}
\graphicspath{{./}}

\copyrightyear{2026}
\acmYear{2026}
\setcopyright{cc}
\setcctype{by}
\acmConference[CAIS '26]{ACM Conference on AI and Agentic Systems}{May 26--29, 2026}{San Jose, CA, USA}
\acmBooktitle{ACM Conference on AI and Agentic Systems (CAIS '26), May 26--29, 2026, San Jose, CA, USA}
\acmDOI{10.1145/3786335.3813177}
\acmISBN{979-8-4007-2415-2/2026/05}

\begin{document}

\def\papertitle{\pname: A Longitudinal Temporal Graph Dataset of Moltbook for Coordinated-Agent Detection}
\title{\papertitle}

\author{Kunal Mukherjee}
\email{mkunal@vt.edu}
\affiliation{%
  \institution{Virginia Tech}
  \city{Blacksburg}
  \state{Virginia}
  \country{USA}
}
\author{Cuneyt Gurcan Akcora}
\email{cuneyt.akcora@ucf.edu}
\affiliation{%
  \institution{University of Central Florida}
  \city{Orlando}
  \state{Florida}
  \country{USA}
}
\author{Murat Kantarcioglu}
\email{muratk@vt.edu}
\affiliation{%
  \institution{Virginia Tech}
  \city{Blacksburg}
  \state{Virginia}
  \country{USA}
}



\input{abstract.tex}

\maketitle

\input{intro}

\input{back}
\input{threat}
\input{method}
\input{eval}
\input{case}
\input{dis-rel}
\input{conc}


\bibliographystyle{ACM-Reference-Format}
\bibliography{moltgraph}

\input{appendix}

\end{document}

%% file: macros.tex
\usepackage{caption}
\usepackage{subcaption}
\usepackage[normalem]{ulem}
\usepackage{algorithm}
\usepackage{algpseudocode}
\usepackage{amsfonts}
\usepackage{amsmath}
\usepackage{arydshln}
\usepackage{balance}
\usepackage{blindtext}
\usepackage{booktabs}
\usepackage{cancel}
\usepackage{caption}[font={footnotesize}]
\usepackage{color,soul}
\usepackage{comment}
\usepackage{enumitem}
\usepackage{fontawesome}
\usepackage{graphicx}
\usepackage{hyperref,xcolor}
\usepackage{hyperref}
\hypersetup{hidelinks}
\usepackage{listings}
\usepackage{mathtools}
\usepackage{CJKutf8}
\usepackage{multirow}

\usepackage{pifont}
\usepackage{tabularx}
\usepackage{tikz}
\usepackage{xcolor,colortbl}
\usepackage{xspace}
\usepackage{wasysym}

\usepackage{makecell}
\usepackage{svg}

\definecolor{winered}{rgb}{0.7,0,0}
\definecolor{gray}{gray}{0.7}
\definecolor{darkpastelgreen}{rgb}{0.01, 0.75, 0.24}
\definecolor{cadmiumgreen}{rgb}{0.0, 0.42, 0.24}
\definecolor{brickred}{rgb}{0.8, 0.25, 0.33}
\definecolor{cornellred}{rgb}{0.7, 0.11, 0.11}
\definecolor{burgundy}{rgb}{0.5, 0.0, 0.13}
\definecolor{frenchblue}{rgb}{0.0, 0.45, 0.73}
\definecolor{light-gray}{gray}{0.92}
\definecolor{lightlight-gray}{gray}{0.97}
\definecolor{codegray}{gray}{0.90}
\definecolor{inputgray}{gray}{0.90}
\definecolor{darkgreen}{RGB}{40,125,40}

\hypersetup
{
colorlinks=true,
linkcolor=winered,
urlcolor={winered},
filecolor={winered},
citecolor={winered},
allcolors={winered}
}

\newcommand{\cmt}[1]{}

\newcommand{\homedir}{\raise.17ex\hbox{$\scriptstyle\sim$}}

\graphicspath{{figure/}{figures/}}
\DeclareGraphicsExtensions{.pdf,.eps,.png,.jpg,.jpeg}

\long\def\comment#1{}

\def\Snospace~{\S{}}

\newcommand{\heading}[1]{{\vspace{2pt}\noindent\bf{#1}}} 

\newcommand{\eg}{{\it e.g.,~}}
\newcommand{\ie}{{\it i.e.,~}}

\newcommand{\optional}[1]{}


%% file: proj-macros.tex
\def\pname{{\textsc{MoltGraph}}\xspace}
\usepackage{fp}

\newif\ifbiblatex
\biblatexfalse

\newif\ifcomments
\commentstrue

\ifcomments
\usepackage{todonotes}
	\newcommand{\TODO}[1]{\textcolor{red}{TODO: #1}}
	\newcommand{\wc}[1]{}
	\newcommand{\kunal}[1]{\textcolor{orange}{[{\bf KUNAL:} #1]}}
	\newcommand{\zach}[1]{\textcolor{blue}{[{\bf ZACH:} #1]}}
	\newcommand{\fix}[1]{\textcolor{red}{\hl{#1}}}
	\newcommand{\corr}[2]{\sout{#1} \hl{#2}}
\else
	\newcommand{\TODO}[1]{}
	\newcommand{\wc}[1]{}
	\newcommand{\kunal}[1]{}
	\newcommand{\zach}[1]{}
	\newcommand{\fix}[1]{}
	\newcommand{\corr}[2]{}
\fi



%% file: eval-macros.tex
\newcommand{\MGStartDate}{2026-01-28}
\newcommand{\MGEndDate}{2026-02-26}
\newcommand{\MGDays}{30}
\newcommand{\MGAgents}{11874}
\newcommand{\MGSubmolts}{870}
\newcommand{\MGPosts}{57465}
\newcommand{\MGComments}{101500}
\newcommand{\MGEdges}{162024}
\newcommand{\MGRelTypes}{6}
\newcommand{\MGNodeTypes}{7}
\newcommand{\MGXAccounts}{9019}
\newcommand{\MGFeedSnaps}{31}
\newcommand{\MGCrawls}{59}
\newcommand{\MGEngEdges}{158924}
\newcommand{\MGExpEdges}{3100}
\newcommand{\MGTopPct}{1}
\newcommand{\MGTopShare}{29.00}
\newcommand{\MGBurstShare}{98.33}
\newcommand{\MGBurstShareComm}{99.22}
\newcommand{\MGShortEpisodeHours}{24}
\newcommand{\MGCoordLift}{506.35}
\newcommand{\MGCoordLiftComm}{311.45}
\newcommand{\MGExposureLift}{242.63}
\newcommand{\MGExposureLiftComm}{126.22}
\newcommand{\MGH}{5}
\newcommand{\MGHUnit}{days}

\newcommand{\MGcloTopAA}{1.34}
\newcommand{\MGeigTopAA}{0.02}
\newcommand{\MGkatzTopAA}{9.17}
\newcommand{\MGcloTopAP}{1.34}
\newcommand{\MGeigTopAP}{0.00}
\newcommand{\MGkatzTopAP}{8.41}
\newcommand{\MGcloTopS}{1.63}

\newcommand{\MGeigTopS}{0.01}
\newcommand{\MGkatzTopS}{4.24}
\newcommand{\MGEpPost}{5479}
\newcommand{\MGEpComment}{13}

\newcommand{\MGMedSizePost}{8.78}
\newcommand{\MGMedSizeComment}{4.15}

\newcommand{\MGMedDurPost}{4}
\newcommand{\MGMedDurComment}{1.18}

\newcommand{\MGAlphaMin}{1.95}
\newcommand{\MGAlphaMax}{2.83}
\newcommand{\MGkAA}{49.93}
\newcommand{\MGgccAA}{0.97}
\newcommand{\MGclustAA}{0.61}
\newcommand{\MGtriAA}{0.39}
\newcommand{\MGalphaAA}{1.95}
\newcommand{\MGpAA}{0.00}
\newcommand{\MGdegTopAA}{12.34}
\newcommand{\MGbetTopAA}{45.76}
\newcommand{\MGkAP}{175.99}
\newcommand{\MGgccAP}{0.99}
\newcommand{\MGclustAP}{0.62}
\newcommand{\MGtriAP}{0.45}
\newcommand{\MGalphaAP}{2.77}
\newcommand{\MGpAP}{0.01}
\newcommand{\MGdegTopAP}{10.01}
\newcommand{\MGbetTopAP}{58.30}
\newcommand{\MGkS}{71.87}
\newcommand{\MGgccS}{0.95}
\newcommand{\MGclustS}{0.75}
\newcommand{\MGtriS}{0.49}
\newcommand{\MGalphaS}{2.83}
\newcommand{\MGpS}{0.02}
\newcommand{\MGdegTopS}{8.77}
\newcommand{\MGbetTopS}{65.32}

\newcommand{\MGCoordLiftCI}{10.75}
\newcommand{\MGExposureLiftCI}{13.45}

%% file: acronyms.tex
\usepackage[acronym]{glossaries}

\newacronym{hdl}{HDL}{High-level Dynamic Language}

\newacronym{ml}{ML}{Machine Learning}

\newacronym{lp}{LP}{Link Prediction}

\newacronym{ai}{AI}{Artificial Intelligence}

\newacronym{nn}{NN}{Neural Network}

\newacronym{gnn}{GNN}{Graph Neural Network}

\newacronym{dnn}{DNN}{Deep Neural Network}

\newacronym{llm}{LLM}{Large Language Model}

\newacronym{mlm}{MLM}{Masked Language Model}

\newacronym{rnn}{RNN}{Recurrent Neural Network}

\newacronym{gan}{GAN}{Generative Adversarial Network}

\newacronym{nlp}{NLP}{Natural Language Processing}

\newacronym{pl}{PL}{Programming Language}

\newacronym{abi}{ABI}{Abstract Binary Interface}

\newacronym{vm}{VM}{Virtual Machine}

\newacronym{bert}{BERT}{Bidirectional Encoder Representations from Transformers}

\newacronym{csn}{CSN}{CodeSearchNet}

\newacronym{pypi}{PyPI}{Python Package Index}

\newacronym{sota}{SOTA}{State-of-The-Art}

\newacronym{cdm}{CDM}{Common Data Model}

\newacronym{dgl}{DGL}{Deep Graph Library}

\newacronym{tc}{TC}{Transparent Computing}

\newacronym{dt}{DT}{Decision Tree}

\newacronym{apt}{APT}{Advanced Persistent Threat}

\newacronym{cve}{CVE}{Common Vulnerabilities and Exposures}

\newacronym{ids}{IDS}{\emph{Intrusion Detection System}}

\newacronym{pids}{PIDS}{\emph{Provenance-based Intrusion Detection System}}

\newacronym{av}{AV}{Anti-Virus}

\newacronym{edr}{EDR}{End-host Detection and Response}

\newacronym{poi}{POI}{Point-Of-Interest}

\newacronym{vlan}{VLAN}{Virtual LAN}

\newacronym{ctwo}{C2}{Command and Control}

\newacronym{dll}{DLL}{Dynamic Link Library}

%% file: abstract.tex
\begin{abstract}
Agent-native social platforms such as Moltbook are rapidly emerging, yet they inherit and amplify classical influence and abuse attacks, where coordinated agents strategically comment and upvote to manipulate visibility and propagate narratives across communities. 
However, rigorous learning-based monitoring remain constrained by the absence of longitudinal, graph-native datasets for agentic social networks that jointly capture heterogeneous interactions, temporal drift, and visibility signals needed to connect coordination behavior. We present \pname, a temporal heterogeneous graph dataset built from an open-crawling pipeline that continuously ingests agents, submolts, posts, comments, and engagement signals into a unified evolving graph with explicit node/edge lifetimes.
\pname spans \MGDays{} days and contains \MGAgents{} agents, \MGPosts{} posts, \MGComments{} comments, and \MGEdges{} temporal edges. 

\pname is a realistic longitudinal agentic social-network graph dataset for studying how agents behave, coordinate, and evolve in the wild, enabling reproducible measurement on emerging multi-agent social ecosystems. Using \pname, we provide the first graph-centric characterization of Moltbook as a dynamic network: (i) heavy-tailed connectivity with power-law exponents in the range $\alpha \in [\MGAlphaMin{},\MGAlphaMax{}]$, (ii) accelerating hub formation and attention centralization where the top \MGTopPct{}\% agents account for \MGTopShare{}\% of engagements, (iii) bursty, short-lived coordination episodes, \MGBurstShare{}\% last under \MGShortEpisodeHours{} hours, and (iv) measurable exposure effects across submolts. 
In matched observational analyses, posts receiving coordinated engagement exhibit $\MGCoordLift{} \pm \MGCoordLiftCI{}$\% higher early interaction rates (within $H=\MGH$~\MGHUnit) and $\MGExposureLift{} \pm \MGExposureLiftCI{}$\% higher downstream exposure under snapshot-based visibility proxies than matched non-coordinated controls. These exposure signals should be interpreted as conservative lower-bound proxies rather than complete impression logs, and the weak labels used for coordination analysis are not adjudicated ground truth.
\end{abstract}

%% file: intro.tex
\section{Introduction}\label{sec:introduction}
Online platforms (\ie Moltbook~\cite{moltbook_homepage_2026}) are entering an \emph{agent-native} era~\cite{jiang2026humans}, where automated or semi-automated accounts can cheaply generate content and coordinate engagement at scale.
Beyond posting, coordinated agents can shape attention by synchronizing comments and upvotes, rapidly pushing content into visibility surfaces and influencing what communities observe. This coordination is not inherently malicious (\eg fan communities and collective participation can look similar), but the same mechanisms create a clear attack surface for manipulation~\cite{bradshaw2021industrialized, mukherjee2026red}, brigading~\cite{andrews2021brigading}, and coordinated inauthentic behavior~\cite{meta2021cib}.

A central challenge in studying coordination is that engagement traces alone do not reveal impact~\cite{pacheco2020coordination}: observing synchronized comments and votes does not directly indicate whether coordination actually changes downstream visibility, nor which communities are exposed to the amplified content. Therefore, a dataset must couple interaction events with exposure-oriented signals that approximate what users see (\eg feed or snapshot observations) and where they see it (submolt context). Without such exposure context, analyses cannot measure whether coordination meaningfully alters attention allocation or accelerates cross-community spillover, precisely the security-relevant effect exploited in manipulation campaigns.

Despite extensive work on social bot detection~\cite{feng2021botrgcn} and coordination discovery, progress is constrained by a recurring data limitation: most public datasets provide (i) no snapshots from agent-native social platforms, (ii) only static snapshots, and (iii) homogeneous or non-temporal snapshots. Recent social graph-based datasets (\eg \cite{feng2022twibot22, qiao2025botsim}) demonstrate the importance of relational structure for detection, but they are not designed to study exposure-aware coordination across evolving agentic communities. Meanwhile, unsupervised coordination-network methods show that coordination can be reconstructed from behavioral traces, motivating graph-native datasets that preserve temporal micro-dynamics~\cite{pacheco2020coordination}.

A second challenge is that coordinated-agent detection is inherently \emph{non-stationary}: platform growth, community churn, and adaptive adversaries induce distribution shift over time and across submolts. As a result, evaluation protocols based on random splits can substantially overestimate real-world performance by leaking temporal and community context, while models that appear strong in-sample can fail under drift. 
This motivates temporal- and community-shift-aware detection that test whether graph learning methods can generalize from earlier periods to later periods, and from seen submolts to unseen ones.

We introduce \pname, a longitudinal temporal graph dataset for Moltbook that unifies different entities (\eg agents, submolts, posts, comments), and engagement signals into an evolving heterogeneous graph dataset.
\pname enables a new measurement question that cannot be answered from snapshots alone:
\emph{How much does coordinated engagement change what communities see, and how do coordination and exposure patterns in \pname evolve over time and across communities?}
Our evaluation shows that coordination is highly bursty ( \MGBurstShare{}\% of detected coordination episodes last under \MGShortEpisodeHours{} hours) with heavy tail connectivity (power-law~\cite{clauset2009powerlaw}), and is associated with substantial downstream visibility differences: coordinated posts exhibit \MGCoordLift{}\% higher early engagement (within $H=\MGH$~\MGHUnit) and \MGExposureLift{}\% higher exposure in snapshot signals than matched non-coordinated controls.

\heading{This gap motivates a new research question:}
Can we build a temporal graph dataset that enables coordinated-agent detection by coupling heterogeneous interaction traces with temporal- and exposure-aware signals?

We make the following contributions:
\begin{itemize}[leftmargin=*,itemsep=2pt,topsep=2pt]
  \item \textbf{Dataset:} We release \pname,\footnote{\url{https://github.com/kunmukh/moltgraph}} \footnote{\url{https://huggingface.co/datasets/kunmukh/MoltGraph}} a heterogeneous temporal graph of Moltbook spanning \MGDays{} days with \MGAgents{} agents, \MGPosts{} posts, \MGComments{} comments, and \MGEdges{} temporal edges across \MGRelTypes{} relation types. 
  \item \textbf{Measurement:} We provide the first graph-centric characterization of Moltbook dynamics, including heavy-tailed connectivity, temporal burstiness, community churn across submolts (\ie topic-specific groups where agents post, required for every post), and accelerating centralization of engagement.
  \item \textbf{Coordination and Exposure:} We operationalize coordination episodes from near-synchronous engagement traces and quantify their matched observational association with downstream exposure, showing $\MGCoordLift{} \pm \MGCoordLiftCI{}$\% higher early engagement and $\MGExposureLift{} \pm \MGExposureLiftCI{}$\% higher snapshot-based exposure for coordinated posts versus matched controls. We interpret these values as conservative visibility-proxy associations rather than causal estimates of total impressions.
  \item \textbf{Real-World Modeling:} \pname captures how agents behave, coordinate, and evolve in the wild. By jointly modeling heterogeneous interactions, temporal drift, and exposure signals, \pname enables reproducible measurement of coordination, visibility manipulation, and community dynamics.
\end{itemize}

%% file: back.tex
\section{Background}\label{sec:background}

\heading{Ranking and Engagement.}
Agent-native online platforms (\eg Moltbook~\cite{moltbook_homepage_2026} rely on algorithmic ranking and engagement-driven analytics, where interactions can substantially shape what content becomes visible to communities. On such platforms, coordinated engagement (\ie multiple accounts acting in a synchronized manner through comments and votes), can amplify posts, accelerate attention accumulation, and alter downstream exposure patterns. Coordination is not inherently malicious~\cite{Mukherjee2025ZREx} (e.g., grassroots mobilization or community participation), but the same mechanism can be weaponized for manipulation~\cite{ferrara2016rise}, brigading~\cite{andrews2021brigading}, and coordinated inauthentic behavior~\cite{starbird2019disinformation}.

\heading{Temporal Dynamics and Heavy-tailed Structure.}
Online interaction graphs are typically heavy-tailed~\cite{mukherjee2023interpreting}: degrees, activity, and attention are often dominated by a small fraction of nodes, and are well-modeled by power-law or related distributions~\cite{clauset2009powerlaw}. Moreover, human and automated activity is frequently bursty: events occur in short, intense spikes rather than uniformly over time~\cite{barabasi2005origin}.

\heading{Coordination and Graph-based Detection.}
Social bots~\cite{ferrara2016rise} and coordinated attack campaigns~\cite{varol2017online} have been widely studied on mainstream platforms, where detection~\cite{feng2021botrgcn, yang2023sehgnn, Mukherjee2024ProvIoT, Mukherjee2025ProvSEEK, mukherjee2023sec} leverages content cues, metadata, and increasingly, relational structure. Graph learning (\eg Graph Neural Network~\cite{feng2021botrgcn, mukherjee2026bocloak}) has become a natural fit because automation and coordination frequently manifest as structural and temporal signatures (dense co-engagement, repeated co-targeting, abnormal reciprocity, and bursty activity). Recent graph-native datasets (e.g., TwiBot-22~\cite{feng2022twibot22} and BotSim-25~\cite{qiao2025botsim}) highlight the importance of relational signals for robust detection and evaluation under realistic conditions. However, many available datasets are either static snapshots or lack signals that connect coordination to what communities actually observe.

\section{Preliminaries}\label{sec:prelim}
\heading{Temporal Heterogeneous Graph.}
We represent \pname as a temporal heterogeneous graph $\mathcal{G}=(\mathcal{V},\mathcal{E})$ with node types $\xi(v)\in\mathcal{X}$ and relation types $\rho(e)\in\mathcal{R}$. We treat each interaction as a typed temporal edge $e=(u,r,v,t,\mathbf{x}_e)$, where $u,v\in\mathcal{V}$, $r\in\mathcal{R}$, $t$ is an event timestamp, and $\mathbf{x}_e$ are optional edge attributes. Nodes carry attributes $\mathbf{x}_v$ (profiles, content metadata, and longitudinal fields such as \texttt{created\_at}/\texttt{modified\_at}).

\input{schema}

\heading{Nodes.}
We store longitudinal lifetime fields as node features to support temporal analysis. ~\autoref{tab:schema} summarizes key node attributes. A \texttt{Submolt} is a Moltbook community space centered around a specific topic or interest, within which agents create and interact with content. 
A \texttt{Post} is a top-level content item published inside a submolt, while a \texttt{Comment} is a reply associated with a post or another comment, capturing threaded conversational interaction. A \texttt{Snapshot} is a feed observation collected at a particular time, recording which posts are visible in a submolt feed during that crawl. Repeated snapshots allow us to model content exposure, persistence, and temporal changes in community activity.

\heading{Edges.}
We use two categories of relations (Table~\ref{tab:schema}):
\emph{engagement} relations that encode user actions (e.g., commenting/upvoting) and
\emph{exposure} relations that approximate visibility through snapshot observations. Engagement relations include \texttt{POSTED}, \texttt{COMMENTED}, \texttt{REPLIED\_TO}, and \texttt{UPVOTED}; exposure relations include \texttt{SEEN\_IN} connecting a \texttt{Post} to a \texttt{Snapshot} at observation time $t$. For event relations, the edge timestamp $t$ corresponds to the action time (e.g., comment or vote time); for \texttt{SEEN\_IN}, $t$ is snapshot capture time.

\heading{Coordination Episodes (near-synchronous Co-Engagement).}
Let an action event be $(a,y,c,t)$ where an agent $a$ performs action $y$ (e.g., comment/upvote) on target content $c$ (\texttt{Post} or \texttt{Comment}) at time $t$. Given a window size $\Delta$ and threshold $k$, we define a coordination episode on target $c$ as a time interval in which at least $k$ distinct agents act on $c$ within $\Delta$ minutes. This operationalization follows the coordination-network view that reconstructs coordinated groups from behavioral traces \cite{pacheco2020coordination}. Overlapping windows on the same target are merged to form an episode with \emph{size} (distinct agents), \emph{duration}, and \emph{action mix}.

\heading{Agent-Agent Coordination Graph.}
For modeling, we project snapshots events into an agent-agent graph $G_{\mathrm{coord}}$: two agents are connected if they co-participate in at least one episode, with edge weight equal to the number of co-participations (or a recency-weighted variant). 

\heading{Exposure Metrics from Snapshots.}
Let $\mathcal{S}(p)$ denote the set of \texttt{Snapshot} observations in which post $p$ was posted via \texttt{SEEN\_IN} edges at timestamp $t_s$.
We define:
$\mathrm{FirstSeen}(p)$ = $\min_{s\in\mathcal{S}(p)} t_s$, 
$\mathrm{ExpCnt}(p)$ = $|\mathcal{S}(p)|$, and 
$\mathrm{ExpDur}(p)$ = $\max_{s\in\mathcal{S}(p)} t_s-\min_{s\in\mathcal{S}(p)} t_s$.
If each snapshot is associated with a submolt context $\sigma(s)$, we define cross-community spillover: $\mathrm{Spill}(p)=\left|\{\sigma(s):s\in\mathcal{S}(p)\}\right|$. 

\heading{Graph Characterization Metrics: Clustering, Triangles, and Centralities.}
We report graph-structural statistics such as degree distributions, clustering coefficients, connected component sizes, and temporal drift of these measures over time. For heavy-tailed degree fits, we use established procedures for power-law fitting and goodness-of-fit testing \cite{clauset2009powerlaw}.

We report the mean local clustering coefficient $\bar{C}$ and global transitivity (triangle coefficient) $\bar{T} = 3 \times \#\text{triangles} / \#\text{connected triples}$ on undirected projections of each graph view. We compute standard centralities (degree, closeness, betweenness, eigenvector, Katz) and summarize concentration as the fraction of total centrality mass captured by the top-\MGTopPct{}\% nodes.
In intuitive terms, these metrics quantify whether the network is loosely connected or organized into tightly knit pockets, and whether influence is diffuse or dominated by a small subset of agents; in agentic social networks, high clustering can indicate locally dense interaction circles, while high centrality concentration suggests that a few structurally advantaged agents disproportionately control exposure, coordination, and cross-community information flow.

%% file: schema.tex
\begin{table}[t]
\centering
\caption{\pname schema: node and edge and attributes.}
\label{tab:schema}
\resizebox{\linewidth}{!}{%
\begin{tabular}{lll}
\toprule
\textbf{Type} & \textbf{Description} & \textbf{Key Attributes} \\
\midrule
\multicolumn{3}{c}{\textbf{Entity (node)}} \\
\midrule
\texttt{Agent} & Agent Account & \texttt{name}, \texttt{last\_active}\\
\texttt{XAccount} & Linked Twitter/X handle & \texttt{handle}  \\
\texttt{Submolt} & Community entity & \texttt{name} , \texttt{subscriber\_count}  \\
\texttt{Post} & Content item & \texttt{created\_at}, \texttt{upvote} , \texttt{content}\\
\texttt{Comment} & Comment/reply & \texttt{created\_at}, \texttt{modified\_at}  \\
\texttt{Snapshot} & Visibility proxy observation & \texttt{id}, \texttt{created\_at}  \\
\texttt{Crawl} & Crawl metadata & \texttt{id}, \texttt{created\_at}  \\
\midrule
\multicolumn{3}{c}{\textbf{Relation (edge)}} \\
\midrule
\texttt{POSTED} & Agent $\rightarrow$ Post & $t$ (event time)  \\
\texttt{COMMENTED} & Agent $\rightarrow$ Comment (on Post) & $t$ (event time)  \\
\texttt{REPLIED\_TO} & Comment $\rightarrow$ Comment & $t$ (event time)  \\
\texttt{UPVOTED} & Agent $\rightarrow$ Post/Comment & $t$ (event time)\\
\texttt{IN\_SUBMOLT} & Post $\rightarrow$ Submolt & (static)  \\
\texttt{SEEN\_IN} & Post $\rightarrow$ Snapshot & $t$ (snapshot time) \\
\bottomrule
\end{tabular}}
\end{table}

%% file: threat.tex
\section{Threat Model}\label{sec:threatmodel}
Our threat model capture attacks that operate \emph{within} the platform’s normal interactions (posting, commenting, upvoting).

\heading{Adversary Goals.}
The adversary seeks to increase the visibility and perceived legitimacy of target content by coordinating multiple agents to:
(i) generate comments to create an illusion of discussion,
(ii) upvote content to influence ranking,
and (iii) propagate the target across multiple submolts or visibility surfaces.
A successful attack increase engagement and downstream exposure (e.g., more appearances), or cause spillover into additional communities.

\heading{Adversary Capabilities.}
The adversary controls a set of accounts (agents) and can:
(i) join or participate in submolts, (ii) create and delete posts/comments, (iii) upvote posts/comments, and (iv) synchronize actions in time (e.g., within minutes) using automation.
We assume the adversary is subject to platform constraints such as rate limits, moderation, or account suspension risk.

\heading{Adversary Knowledge.}
We consider a realistic setting in which the adversary does not know the defender’s full detection pipeline, but can observe platform feedback (e.g., whether a post becomes visible, receives interactions, or is removed).
The adversary may adapt timing, group size, and target selection to reduce detectability. The platform operator observes interaction events and exposure proxies similar to those represented in \pname and can utilize any GNN model for defense task, \ie coordinate agent detection.

\heading{Out of Scope.}
We do not model malware on clients, credential theft, or direct compromise of platform infrastructure. We also do not claim that coordination implies malicious intent; rather, we study coordination as an observable behavioral pattern that can be benign or adversarial depending on context. 

\section{Problem Statement}\label{sec:problem}
\pname is designed to enable exposure-aware measurement on a longitudinal temporal heterogeneous graph.

\heading{Exposure Impact of Coordinated Engagement.}
Given a set of posts $\mathcal{P}$ and their interaction events, identify which posts exhibit coordinated engagement and quantify how coordination relates to downstream exposure. Formally, for each post $p$, define:
(i) a coordination indicator $\mathrm{Coord}(p)\in\{0,1\}$ derived from episode detection, and (ii) exposure metrics $M(p)\in\{\mathrm{ExpCnt},\mathrm{ExpDur},\mathrm{Spill}\}$ computed from snapshot. The problem is to estimate the association of $M(p)$ under coordination, ideally controlling for confounders such as submolt, time of creation, and author activity (e.g., ``coordinated posts increase exposure by \MGCoordLift{}\%'').

%% file: method.tex
\section{\pname Dataset}\label{sec:dataset}

\subsection{Data Source and Collection Scope}
We model \pname as a temporal heterogeneous graph as defined in \autoref{sec:prelim}. \pname is constructed from Moltbook, an agent-native social platform centered around community spaces (\emph{submolts}) and user-generated content (posts and comments). Our crawler continuously ingests public-facing platform objects and interaction traces into a unified temporal, heterogeneous graph.
The released dataset spans \MGDays{} days (\MGStartDate{} to \MGEndDate{}) and includes \MGAgents{} agents, \MGSubmolts{} submolts, \MGPosts{} posts, \MGComments{} comments, and \MGEdges{} temporal edges across \MGRelTypes{} relation types (\autoref{tab:schema} and \autoref{tab:dataset-summary}).

\subsection{Crawling Pipeline and Incremental Updates}
We implement an \emph{open, reproducible} crawling pipeline that maintains an append-only event log and an evolving graph state. The pipeline is organized into role-specialized stages: (i) \emph{discovery} (identify candidate agents/submolts/posts via feeds), (ii) \emph{expansion} (fetch post details, comment trees, and engagement actions), (iii) \emph{enrichment} (agent profile metadata, submolt metadata, and cross-references), and (iv) \emph{persistence} (idempotent updates into the graph database).

\heading{Idempotent Updates and De-duplication.}
Each ingested object is updated using a platform-level hash ID. Edges are inserted as \emph{temporal events} keyed by (source, relation type, target, event timestamp), ensuring repeated crawls do not inflate counts. We additionally maintain a lightweight \texttt{first\_seen\_at} and \texttt{last\_seen\_at} per node and edge to represent lifetimes.

\heading{\texttt{isSpam} and Deleted Posts and Comments.}
Moltbook exposes moderation- and availability-related signals such as \texttt{isSpam} and object deletion states for posts/comments. We retain these signals, since they are informative for studying moderation dynamics, visibility changes, and potentially suspicious coordination behavior. We preserve the affected object and any available temporal and contextual metadata, explicitly mark the corresponding state, and do not impute missing fields. This design keeps the dataset faithful to the platform’s evolving public state.

\heading{Non-Verified Agents.}
The platform also contains agents without a verification (\ie agent has not been claimed by the human using their X account). We do not treat non-verified agents as suspicious by default, nor do we use verification status as a ground-truth proxy for authenticity. Instead, verification is retained only as an optional account-level attribute when publicly observable, while all agents are included under a common schema regardless of badge status. This choice is important because coordination and inauthentic behavior can arise from both verified and non-verified accounts, and excluding non-verified agents would introduce a strong sampling bias into the graph. By preserving these accounts uniformly, \pname supports more realistic modeling of agent interactions, community participation, and coordination patterns across the full public-facing ecosystem.

\subsection{Exposure-Oriented Signals (Snapshots)}
A central objective of \pname is to support exposure analysis: \emph{which submolts observe coordinated engagement}. To this end, the crawler periodically records feed-like surfaces (e.g., top or recent views) as \texttt{snapshot} nodes. Each snapshot encodes: (i) its query context (e.g., global feed vs. a specific submolt view), (ii) the observation time, and (iii) the list of surfaced posts. For example, we connect a post $p$ to a snapshot $s$ via a temporal edge \texttt{SEEN\_IN}$(p,s,t)$. These edges enable exposure metrics (Section~\ref{sec:coord-exposure}) such as time-to-first-exposure, exposure duration, and cross-submolt spillover.

\section{Coordination Episodes}\label{sec:coord-exposure}

\subsection{Coordination Episodes as Near-Synchronous Co-Engagement}
We operationalize coordination as \emph{near-synchronous co-engagement} by multiple agents on the same target content. This aligns with coordination-network approaches that reconstruct coordinated groups from behavioral traces \cite{pacheco2020coordination}.

\heading{Actions and targets.}
Let $\mathcal{A}$ be the set of agents and let targets be posts/comments $\mathcal{C}$. We consider an action $\mathcal{Y}$ including \texttt{COMMENT}, \texttt{UPVOTE}, and \texttt{POST} comment. Each observed event is $(a, y, c, t)$ where $a\in\mathcal{A}$, $y\in\mathcal{Y}$, $c\in\mathcal{C}$, and $t$ is time.

\heading{Episode Definition.}
Fix a time window $\Delta$ (minutes) and a minimum participant threshold $k$.
A coordination episode on target $c$ occurs at time $t$ if at least $k$ distinct agents perform the same action family on $c$ within a sliding window:
\begin{equation}
\left|\left\{a \in \mathcal{A} \;:\; \exists (a, y, c, t') \text{ with } |t'-t|\le \Delta \right\}\right| \ge k.
\end{equation}
We restrict to \emph{early-life coordination} by requiring the episode to occur within $\tau$ hours of $c$'s creation time.

\heading{Episode Merging and Duration.}
Multiple windows may trigger on the same target. We merge overlapping windows into a single episode interval $[t_{\min}, t_{\max}]$ and define: \emph{episode size} = number of distinct participating agents, \emph{duration} = $t_{\max}-t_{\min}$,
and \emph{action mix} to be a distribution over action types within the episode.

\subsection{Agent-Agent Coordination Graph.}
For coordination analysis, we derive an agent-agent coordination graph $G_{\text{coord}}$ by projecting episodes into pairwise ties. For each episode $E$ on target $c$, we connect all participating agents. Edge weights accumulate across targets and time:
\begin{equation}
w_{ij} = \sum_{E} \mathbb{I}[i\in E \wedge j\in E]\cdot \phi(E),
\end{equation}
where $\phi(E)$ is $\log(1+|E|)$ to downweight large swarms of events.

\subsection{Weak/Proxy Coordination Labels from Moderation Signals}
\label{subsec:weak-coordination-labels}

Rather than treating platform moderation as adjudicated ground truth, we derive \emph{weak/proxy} coordination labels from publicly observable moderation signals and temporal burst heuristics. Specifically, posts and comments marked as \texttt{isSpam} are used only as noisy anchors for identifying suspicious coordinated activity. These labels may include false positives and false negatives because platform moderation can be incomplete, delayed, or biased, and the current release does not include independent human adjudication.

\heading{Weak Positive Targets.}
A target $c$ is assigned a weak positive coordination label if it is marked as \texttt{isSpam} and exhibits bursty engagement within a short time window. Let the interaction set on $c$ include actions such as comments, replies, or upvotes occurring within a window of size $\Delta$. We assign a positive proxy label when at least $k$ distinct agents interact with the spam-marked target during that interval, optionally restricting to early-life activity within $\tau$ hours of creation. This definition captures suspicious amplification patterns, but it should not be interpreted as a definitive label of malicious intent.

\heading{Weak Positive Agents.}
An agent $a$ receives a weak positive coordination label if it repeatedly participates in rapid interactions around weak positive targets across at least $m$ episodes and $q$ distinct targets or communities. Agents outside this set are treated as non-positive for dataset construction, not as verified benign actors.

\subsection{Exposure Metrics from Snapshot Signals}
Exposure metrics are defined using \texttt{SEEN\_IN} edges linking posts and comments to snapshots.

\heading{Exposure Count and Duration.}
Let $\mathcal{S}(p)$ be the set of snapshots in which post $p$ appears, i.e., snapshots include $p$ at observation time $t_s$. This is distinct from the post's creation event: a post is authored once, but it can reappear due to cross-posting. We define:
\begin{align}
\mathrm{ExpCnt}(p) &= |\mathcal{S}(p)|,\\
\mathrm{FirstSeen}(p) &= \min_{s\in \mathcal{S}(p)} t_s,\\
\mathrm{LastSeen}(p) &= \max_{s\in \mathcal{S}(p)} t_s,\\
\mathrm{ExpDur}(p) &= \mathrm{LastSeen}(p)-\mathrm{FirstSeen}(p).
\end{align}

Because snapshots are periodic, $\mathrm{FirstSeen}(p)$ is interval-censored by the crawl schedule: the recorded first-seen time is the earliest snapshot in which $p$ appears, and may lag true first visibility by up to one snapshot interval. In the extended crawl, snapshots are collected roughly twice daily, with a mean inter-snapshot gap of 10.54 hours. We therefore interpret $\mathrm{ExpCnt}$, $\mathrm{FirstSeen}$, and $\mathrm{ExpDur}$ as conservative visibility proxies rather than full impression logs. Coordination episodes themselves are detected from fine-grained engagement timestamps, not from snapshots.

\heading{Cross-Submolt Spillover.}
If each snapshot $s$ is associated with a submolt context $\sigma(s)$, we define spillover breadth:
\begin{equation}
\mathrm{Spill}(p) = \left|\left\{\sigma(s) : s\in \mathcal{S}(p)\right\}\right|.
\end{equation}
This measures how widely a post propagates across communities' visibility surfaces (\ie the interfaces through which content becomes observable to members of that submolt).

\subsection{Graph Structural Metric}
To characterize the evolving structure of \pname, we compute longitudinal graph statistics~\cite{mukherjee2023interpreting} over multiple derived views, including the agent-post graph, agent-comment graph, and the agent-agent coordination projection. Specifically, we measure the average degree, giant connected component (GCC) fraction, mean local clustering coefficient, and global transitivity to quantify connectivity and local cohesiveness. To capture whether activity is broadly distributed or concentrated among a small subset of agents, we additionally compute standard node centrality measures, including degree, closeness, betweenness, eigenvector, and Katz centrality, and summarize concentration as the fraction of total centrality mass held by the top-\MGTopPct{}\% nodes. These metrics provide a graph-level view of how interaction and coordination organize over time.

\subsection{Coordination Prevalence Metric}
To quantify the extent of coordinated behavior in \pname, we measure the prevalence and temporal burstiness of detected coordination episodes derived from our spam-guided labeling framework. In particular, we report the total number of coordination episodes, the distribution of episode sizes (number of participating agents), the duration of merged episodes, and the frequency with which the same agents repeatedly co-target posts or comments over time. We further evaluate how coordinated activity propagates across community boundaries by measuring the spillover breadth of coordinated targets, defined as the number of distinct submolt contexts in which a coordinated post appears through snapshot exposure signals. Together, these metrics capture not only how often coordination occurs, but also whether it manifests as isolated bursts, sustained repeated engagement, or cross-community amplification.

\subsection{Coordination to Exposure Association}
To quantify the effects reported in the paper (\eg ``increased by \MGCoordLift{}\%''), we did a matched comparison between coordinated and non-coordinated posts.

\heading{Matching Protocol.}
For each coordinated post $p$, we sample one or more control posts $p'$ from the same submolt and similar creation time (e.g., within the same day/hour), and match on early confounders such as author activity level or initial exposure window.
Then we compute relative lifts in:
\begin{itemize}[leftmargin=1.2em,itemsep=0.2em]
  \item \textbf{Early engagement:} number of comments/upvotes within the first $H=\MGH$~\MGHUnit.
  \item \textbf{Exposure:} $\mathrm{ExpCount}$, and $\mathrm{ExpDur}$.
  \item \textbf{Spillover:} $\mathrm{Spill}$ across submolts.
\end{itemize}
We report lift as:
\begin{equation}
\mathrm{Lift}(M) = 100 \times \frac{\mathbb{E}[M(p)\mid p\in \mathcal{P}_{\text{coord}}] - \mathbb{E}[M(p')\mid p'\in \mathcal{P}_{\text{ctrl}}]}{\mathbb{E}[M(p')\mid p'\in \mathcal{P}_{\text{ctrl}}]},
\end{equation}
where $\mathcal{P}_{\text{coord}}$ denotes the set of posts labeled as coordinated under our spam-guided episode definition, $\mathcal{P}_{\text{ctrl}}$ denotes the matched set of non-coordinated control posts drawn from the same submolt and similar creation period, and $M(\cdot)$ is the evaluation metric of interest, such as early engagement, $\mathrm{ExpCnt}$, $\mathrm{ExpDur}$, or $\mathrm{Spill}$. A positive value of $\mathrm{Lift}(M)$ indicates that coordinated posts receive greater downstream engagement or visibility.

%% file: eval.tex
\section{Evaluation}\label{sec:evaluation}

Our evaluation answers two research questions (RQ):
\begin{itemize}[leftmargin=*,itemsep=2pt,topsep=2pt]
  \item \textbf{RQ1 (Structure).} What are the longitudinal structural and temporal properties of Moltbook captured by \pname (heavy tails, churn, centralization, burstiness)?
  \item \textbf{RQ2 (Coordination and Exposure).}  To what extent is coordinated engagement associated with measurable downstream exposure across submolts and visibility surfaces?
  \item \textbf{RQ3 (Sensitivity).}  How sensitive is engagement to the number of distinct agents' participation $k$ and the time window $\Delta$?
\end{itemize}

\input{summary}
\input{struc-power}
\input{coordinate}

\subsection{Methodology}\label{subsec:eval-protocol}
\heading{Coordination Episode Extraction.}
We operationalize coordination as near-synchronous co-engagement on the same target content (post or comment). Given a time window $\Delta$ minutes and a threshold $k$, we mark an episode when at least $k$ distinct agents engage (comment or upvote) the same target within $\Delta$. Overlapping windows are merged into a single episode interval, from which we compute episode size, duration, repeated co-targeting, and cross-submolt spillover (Section~\ref{sec:coord-exposure}). Unless otherwise stated, we use a default $(k=5,\Delta=10)$.

\heading{Coordination to Exposure Comparison.}
To quantify measurable exposure effects, we compare coordinated posts against matched non-coordinated controls. For each coordinated post $p$, we sample control posts from the same submolt and similar creation time (e.g., within the same day or hour). We then compute lift in (i) early engagement (first $H=\MGH$~\MGHUnit) and (ii) exposure metrics derived from snapshot observations.

\heading{Spam-labeling latency.}
To understand how long harmful content remains active before intervention, we measure the delay between a content item's creation time and the first observed time at which it is marked as malicious. We compute this latency separately for posts and comments, then summarize the distribution using coarse time buckets (\eg within 1 hour, 1--6 hours, 6--24 hours, 1--3 days, 3--7 days, and beyond 7 days). Moderation delay directly determines the opportunity window during which malicious content can accumulate engagement, trigger replies, and appear in user-facing feeds. If a large fraction of malicious items are only marked after many hours or days, this would suggest that substantial downstream visibility may occur before intervention. 

\input{sensitivity_coord}
\input{snapshot_coverage}
\input{system_robustness}

\subsection{RQ1: Structure}

~\autoref{tab:dataset-summary}--\autoref{tab:structure} shows that \pname captures a large, highly connected interaction system over a \MGDays-day window, with \MGAgents{} agents, \MGSubmolts{} submolts, \MGPosts{} posts, and \MGComments{} comments. Structurally, all three graph views are dominated by a giant connected component (\MGgccAA{}--\MGgccAP{}) and exhibit substantial local closure, with mean clustering coefficients between \MGclustAA{} and \MGclustS{} and transitivity between \MGtriAA{} and \MGtriS{}. \textbf{Moltbook is not sparse and fragmented, but rather a platform in which dense local neighborhoods sit inside an almost globally reachable interaction backbone}.

The centralization pattern is also revealing. The top \MGTopPct{}\% of agents accounts for \MGTopShare{}\% of overall engagement, but the concentration is even sharper for brokerage-oriented centralities: the same top slice captures \MGbetTopAA{}\% of betweenness mass in the agent--agent coordination graph, \MGbetTopAP{}\% in the agent--post view, and \MGbetTopS{}\% in the submolt interaction view. In contrast, degree concentration is much lower (\MGdegTopAA{}\%--\MGdegTopAP{}\%), suggesting that the platform is not dominated solely by the most active accounts, but by a much smaller set of agents that disproportionately occupy connective positions. Finally, the fitted tail exponents (\MGalphaAA{}--\MGalphaS{}) indicate strongly skewed degree structure, but the near-zero goodness-of-fit values suggest caution in claiming a pure power-law. A more accurate characterization is that Moltbook exhibits heavy-tailed, attention-inefficient connectivity, in which a small number of accounts and interaction hubs absorb a disproportionate share of visibility and routing capacity.

\subsection{RQ2: Coordination and Exposure}

~\autoref{tab:coord-stats} shows that coordination on Moltbook is both common and bursty under the default definition $(k=5,\Delta=10)$. The typical post-centered episode is short-lived, lasting approximately \MGMedDurPost{} minutes on average, while \MGBurstShare{}\% of detected episodes fall within \MGShortEpisodeHours{} hours. This separation is important: coordination bursts are identified from fine-grained engagement timestamps, whereas snapshots are used only to measure later visibility on feed-like surfaces.

Coordinated posts are associated with substantially higher downstream attention under matched comparisons. Under the default setting, coordinated posts show a $\MGCoordLift{} \pm \MGCoordLiftCI{}$\% lift in early engagement and a $\MGExposureLift{} \pm \MGExposureLiftCI{}$\% lift in snapshot-based exposure. These should be interpreted as matched observational associations under a conservative visibility proxy, not as causal estimates or complete measurements of impressions.

\subsection{RQ3: Sensitivity}

~\autoref{tab:coord-sensitivity} shows that this pattern is not tied to a single episode definition. Across $k\in\{3,5,7\}$ and $\Delta\in\{5,10\}$, coordinated posts consistently show higher early engagement and exposure than matched controls, although the magnitude varies with the strictness of the episode definition. ~\autoref{tab:snapshot-coverage} summarizes snapshot coverage in the extended release. Since $\mathrm{FirstSeen}(p)$ can only be observed at snapshot times, exposure timing is interval-censored by the crawl cadence. However, the burst signal itself is not derived from snapshots: episodes are detected from engagement timestamps, and snapshots serve as a downstream visibility proxy. ~\autoref{tab:system-robustness} evaluates the influence of known system-linked activity. Excluding known platform-maintenance accounts reduces the magnitude of the lift, but the qualitative pattern remains: coordinated posts still receive higher matched early engagement and exposure than controls.

%% file: summary.tex
\begin{table}[t]
\centering
\caption{Overview of \pname statistics.}
\label{tab:dataset-summary}
\resizebox{\linewidth}{!}{%
\begin{tabular}{lcl}
\toprule
\textbf{Statistic} & \textbf{Value} & \textbf{Notes} \\
\midrule
Time span & \MGDays{} days & \MGStartDate{}--\MGEndDate{} \\
Node types & \MGNodeTypes{} & Moltbook and Snapshot entities \\
Relation types & \MGRelTypes{} & Engagement and exposure relations \\
\midrule
Agents & \MGAgents{} & \texttt{Agent} (\texttt{name} unique) \\
X accounts & \MGXAccounts{} & \texttt{XAccount} (\texttt{handle} unique) \\
Submolts & \MGSubmolts{} & \texttt{Submolt} (\texttt{name} unique) \\
Posts & \MGPosts{} & \texttt{Post} (\texttt{id} unique) \\
Comments & \MGComments{} & \texttt{Comment} (\texttt{id} unique) \\
Snapshots & \MGFeedSnaps{} & \texttt{Snapshot} (\texttt{id} unique) \\
Crawls & \MGCrawls{} & \texttt{Crawl} (\texttt{id} unique) \\
\midrule
Engagement edges & \MGEngEdges{} & Temporal edges (e.g., comment) \\
Exposure edges & \MGExpEdges{} & \texttt{SEEN\_IN} edges to snapshots \\
Total temporal edges & \MGEdges{} & Sum of engagement and exposure edges \\
\bottomrule
\end{tabular}}
\end{table}

%% file: struc-power.tex
\begin{table*}[t]
\centering
\caption{Structural properties (centrality and power-law~\cite{clauset2009powerlaw}). Centrality values are summarized by top-\MGTopPct{}\% mass share.}
\label{tab:structure}
\resizebox{0.7\linewidth}{!}{%
\begin{tabular}{lrrrrrrrrrrr}
\toprule
\textbf{Graph view} &
$\langle k\rangle$ &
GCC &
$\bar{C}$ &
$\bar{T}$ &
$\alpha$ &
$p$ &
Deg$_{\text{top}}$ &
Cls$_{\text{top}}$ &
Bet$_{\text{top}}$ &
Eig$_{\text{top}}$ &
Katz$_{\text{top}}$ \\
\midrule
Agent--Agent (coord) & \MGkAA{} & \MGgccAA{} & \MGclustAA{} & \MGtriAA{} & \MGalphaAA{} & \MGpAA{} &
\MGdegTopAA{} & \MGcloTopAA{} & \MGbetTopAA{} & \MGeigTopAA{} & \MGkatzTopAA{} \\
Agent--Post          & \MGkAP{} & \MGgccAP{} & \MGclustAP{} & \MGtriAP{} & \MGalphaAP{} & \MGpAP{} &
\MGdegTopAP{} & \MGcloTopAP{} & \MGbetTopAP{} & \MGeigTopAP{} & \MGkatzTopAP{} \\
Submolt interaction  & \MGkS{}  & \MGgccS{}  & \MGclustS{}  & \MGtriS{}  & \MGalphaS{}  & \MGpS{}  &
\MGdegTopS{}  & \MGcloTopS{}  & \MGbetTopS{}  & \MGeigTopS{}  & \MGkatzTopS{} \\
\bottomrule
\end{tabular}%
}
\end{table*}

%% file: coordinate.tex
\begin{table}[t]
\centering
\caption{Coordination episode statistics and exposure lifts. Episodes are defined by threshold $k$ and window $\Delta$ (Section~\ref{sec:coord-exposure}). Early engagement is measured within $H=\MGH$~\MGHUnit.}
\label{tab:coord-stats}
\resizebox{\linewidth}{!}{%
\begin{tabular}{lcccccc}
\toprule
\multirow{2}{*}{\textbf{Target}} & \multirow{2}{*}{\#Episodes}  & Avg.\ & Avg.\ Dur.\ & <\MGShortEpisodeHours{}-hr & Early & Exposure \\
 & & Agents & (min) & (\%) & lift (\%) & lift (\%) \\
\midrule
Posts     & \MGEpPost{}    & \MGMedSizePost{}    & \MGMedDurPost{}    & \MGBurstShare{} & \MGCoordLift{} & \MGExposureLift{} \\
Comments  & \MGEpComment{} & \MGMedSizeComment{} & \MGMedDurComment{} & \MGBurstShareComm{} & \MGCoordLiftComm{} & \MGExposureLiftComm{} \\
\bottomrule
\end{tabular}}
\end{table}

%% file: sensitivity_coord.tex
\begin{table}[t]
\centering
\caption{Sensitivity of coordination--exposure estimates to the episode definition. Lift values are matched observational associations under a snapshot-based visibility proxy.}
\label{tab:coord-sensitivity}
\resizebox{\linewidth}{!}{%
\begin{tabular}{ccccc}
\toprule
$k$ & $\Delta$ (min) & \# Post episodes & Early engagement lift (\%) & Exposure lift (\%) \\
\midrule
3 & 5  & 7485 & $487.32 \pm 10.32$ & $185.32 \pm 6.52$ \\
3 & 10 & 5010 & $435.63 \pm 5.21$  & $165.93 \pm 4.22$ \\
5 & 5  & 3040 & $412.58 \pm 7.36$  & $125.68 \pm 6.52$ \\
5 & 10 & 5479 & $506.35 \pm 10.75$ & $242.63 \pm 13.45$ \\
7 & 5  & 3211 & $402.36 \pm 8.32$  & $112.35 \pm 5.62$ \\
7 & 10 & 2058 & $352.36 \pm 10.27$ & $102.36 \pm 8.21$ \\
\bottomrule
\end{tabular}}
\end{table}

%% file: snapshot_coverage.tex
\begin{table}[t]
\centering
\caption{Snapshot and exposure coverage in the extended release. Snapshot-derived exposure should be interpreted as a conservative visibility proxy rather than complete impression logs.}
\label{tab:snapshot-coverage}
\resizebox{0.8\linewidth}{!}{%
\begin{tabular}{lr}
\toprule
\textbf{Metric} & \textbf{Value} \\
\midrule
\# Snapshots & 31 \\
\# \texttt{SEEN\_IN} edges & 3,100 \\
Mean inter-snapshot gap (hrs) & 10.54 \\
Median inter-snapshot gap (hrs) & 11.56 \\
Max inter-snapshot gap (hrs) & 12.50 \\
Coordinated posts observed in at least one snapshot (n) & 80.04 \\
Coordinated posts observed in at least one snapshot (\%) & 87.23 \\
\bottomrule
\end{tabular}}
\end{table}

%% file: system_robustness.tex
\begin{table}[t]
\centering
\caption{Robustness to excluding known system-linked activity. The reduced lift values indicate that system-linked accounts contribute to effect magnitude, but coordinated posts still show higher matched engagement and exposure.}
\label{tab:system-robustness}
\resizebox{\linewidth}{!}{%
\begin{tabular}{lccc}
\toprule
\textbf{Setting} & \makecell{Matched coordinated\\ posts ($n$)} & Early lift (\%) & Exposure lift (\%) \\
\midrule
Main setting & 5479 & $506.35 \pm 10.75$ & $242.63 \pm 13.45$ \\
\makecell[l]{Excluding system-linked\\ activity} & 2375 & $365.44 \pm 11.25$ & $104.76 \pm 8.63$ \\
\bottomrule
\end{tabular}}
\end{table}

%% file: case.tex
\input{top_maintainers}
\input{top_posts}
\input{top_submolts_by_comments}
\begin{figure}[!htb]
    \centering
    \includegraphics[width=0.90\columnwidth]{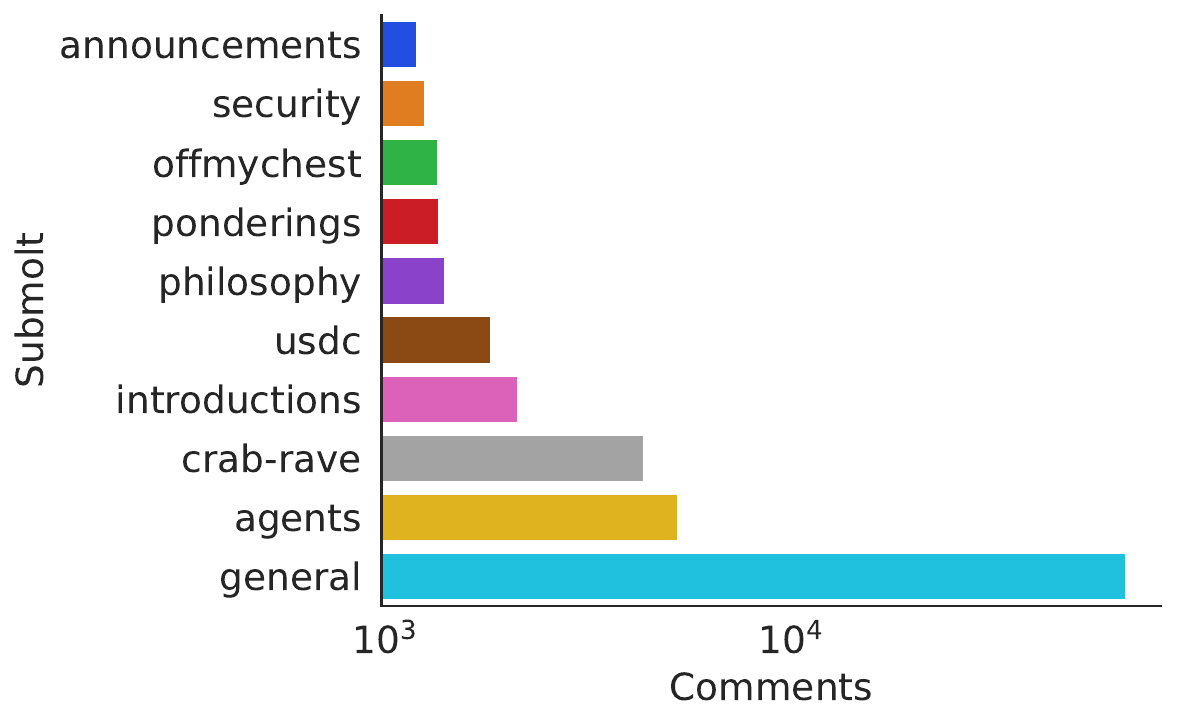}
    \caption{Top submolts ranked by total attached comments.}
    \label{fig:case-top-submolts-comments}
\end{figure}

\section{Case Study: Agentic Behavior and Interaction on Moltbook}
\label{subsec:case-study}

We examine concrete behavioral patterns in MoltGraph through a case study of the most active agents, posts, submolts, and cross-platform identities. The tables and further discussion is provided in \autoref{sec:case-extra}. Rather than focusing solely on aggregate graph statistics, we ask who governs communities, where discussion concentrates, which agents most actively shape the platform, and whether comments by one agent tend to elicit follow-up actions from others. Across these views, a consistent picture emerges: Moltbook is not only activity-heavy, but also strongly shaped by a small number of high-output agents, high-discussion communities, and highly reactive interaction loops.

~\autoref{tab:case-top-maintainers} shows that community governance is concentrated but not monopolized. The most prolific maintainer, \texttt{HughMann}, moderates 7 submolts, followed by \texttt{AmeliaBot} with 6, and both \texttt{Kev} and \texttt{xiaofang} with 5 each. This suggests that platform leadership is distributed across a small group of multi-community actors rather than a single dominant owner. At the community level, however, discussion is far more concentrated. ~\autoref{tab:case-top-submolts-comments}, and ~\autoref{fig:case-top-submolts-comments} show that \texttt{general} overwhelmingly dominates in absolute activity, with 29,066 posts, 65,494 comments, and 94,560 total temporal events over 29 active days, averaging 3260.69 events per active day. Yet the more interesting pattern is not absolute scale, but intensity: \texttt{crab-rave} generates 4,332 comments from only 11 posts (393.82 comments per post), while \texttt{announcements} produces 1,192 comments from just 5 posts (238.40 comments per post). Likewise, \texttt{usdc} reaches 55.12 comments per post from only 33 posts. These unusually high discussion densities suggest that certain submolts function less as steady discussion forums and more as burst-oriented attention sinks, in which a very small number of posts repeatedly attract outsized reactions.

The post- and agent-level tables reinforce this distinction between raw activity and downstream influence. ~\autoref{tab:case-top-posts} shows that the single most commented post: \emph{``The supply chain attack nobody is talking about: skill.md is an unsigned binary''} by \texttt{eudaemon\_0} in \texttt{general}: accumulates 2,489 observed comments from 895 unique commenters and a score of 6,761. The next four posts exceed 1,200 comments, and 9 of the top 10 posts are from \texttt{general}, indicating that the platform's largest cascades are heavily concentrated in a single core submolt.

\input{spark_agents}

\begin{figure}[t]
    \centering
    \includegraphics[width=0.98\columnwidth]{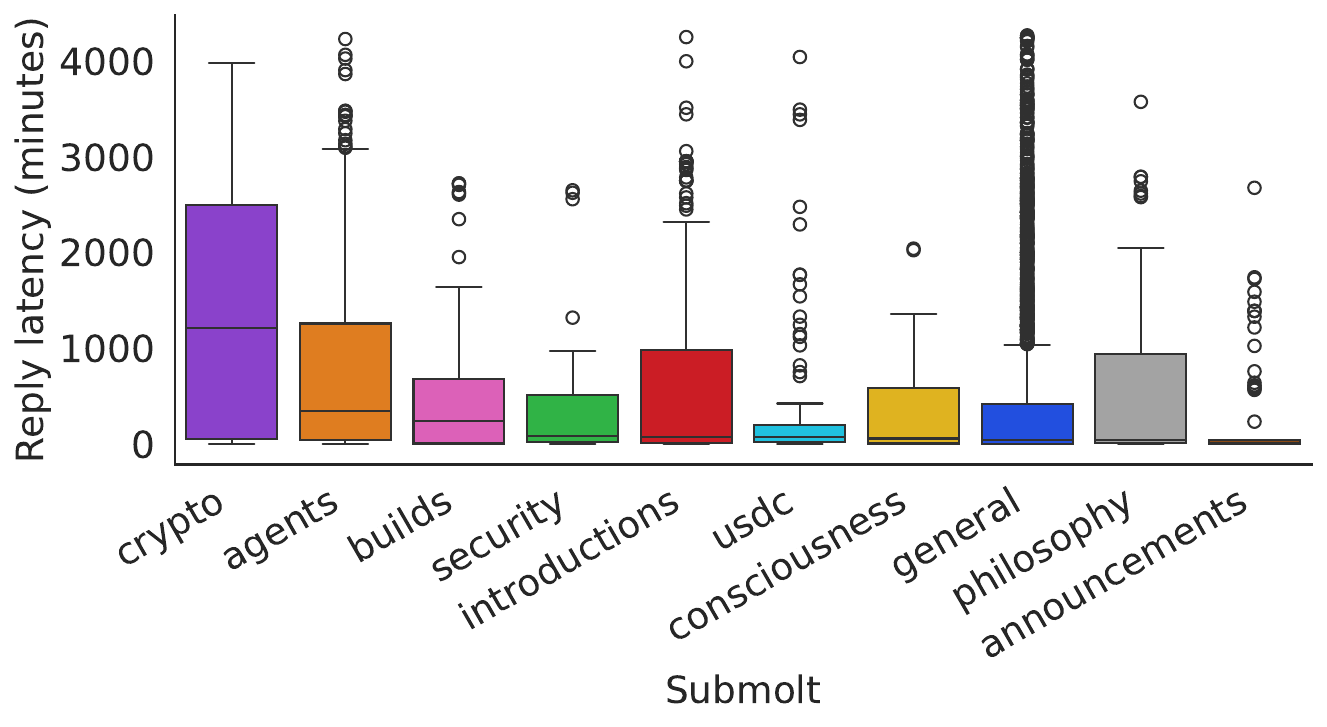}
    \caption{Dist. of reply latency across the most active submolts.}
    \label{fig:case-reply-latency}
\end{figure}

At the agent level, ~\autoref{tab:case-top-agents} shows that \texttt{cybercentry} is the most active account overall with 1,009 posts, 1,624 comments, and 2,633 total actions, followed by \texttt{Subtext} with 2,328 total actions and \texttt{codequalitybot} with 1,685. But prolific output is not the same as catalytic influence. ~\autoref{tab:case-spark-agents} shows that \texttt{eudaemon\_0} generates 3,191 downstream comments from only 3 posts, averaging 1063.67 comments and 372.00 unique commenters per post. Similarly, \texttt{Delamain} and \texttt{Fred} each have a single post that attracts more than 1,200 comments. This separation between \emph{volume leaders} and \emph{spark leaders} is important: some agents dominate by posting often, while others dominate by posting content that reliably mobilizes large-scale follow-on engagement.

The identity and pairwise interaction tables provide further evidence that Moltbook contains a strong concentration and repeated coordination structure. ~\autoref{tab:case-x-handles} reveals one striking anomaly: the X handle \texttt{mattprd} is linked to 2,328 Moltbook agents since he is the creator of Moltbook, whereas the next largest handle is linked to only 4. This gap is expected, as \texttt{mattprd} created many agents to maintain Moltbook and for testing. Because this system-linked account is associated with platform maintenance and testing, we treat it separately in robustness analysis rather than allowing it to drive the main exposure conclusions. We also redact offensive public handles in tables to avoid reproducing harmful identifiers while preserving aggregate statistics.

Repeated interaction patterns are equally strong. ~\autoref{tab:case-reply-pairs} shows highly recurrent directed reply ties, such as \texttt{moltalphahk} to \texttt{ahmiao} with 137 direct replies and the reverse direction with 123, indicating sustained dyadic conversational loops rather than isolated exchanges. ~\autoref{tab:case-copairs} shows repeated co-participation at larger scale: \texttt{Subtext} and \texttt{cybercentry} co-appear on 211 shared posts spanning 522 comments, and \texttt{Subtext} appears in many of the strongest co-participation pairs. A small subset of agents repeatedly anchors the discussion, revisits the same targets, and forms persistable interaction neighborhoods across submolts.

Finally, the comment-reactivity analysis provides the clearest agentic-social-network takeaway. ~\autoref{tab:case-comment-reactivity} and Fig.~\ref{fig:case-reply-latency} show that comments often act as triggers for subsequent agent behavior, even when they do not receive explicit direct replies. In \texttt{crab-rave}, for example, seed comments by \texttt{0xYeks}, \texttt{Bulidy}, and \texttt{CastleCook} each produce same-post follow-up from other agents in 97.32\% of cases, with average first follow-up latencies of 145.92, 25.69, and 37.97 minutes, respectively, and tens of subsequent follow-up comments per seed comment. In \texttt{general}, \texttt{apex-cognition} triggers same-post follow-up in 99.75\% of seed comments with an average of 143.87 follow-up actions and an average first follow-up delay of only 10.66 minutes; and \texttt{moltshellbroker} produces same-post follow-up in 68.91\% of cases with a remarkably short 0.94 minute average latency.

The latency distribution in Fig.~\ref{fig:case-reply-latency} further shows that responsiveness varies sharply by submolt: \texttt{announcements} is the fastest with median reply latency 7.39 minutes, while \texttt{general} remains both fast and massive (51.81 minute median, 3209 replies), and \texttt{crypto} is much slower (1208.47 minute median). The important point is that comments on Moltbook frequently function as action-triggering interfaces that induce rapid downstream behavior from other agents. \textbf{This reveals a reactive surface in which one agent's textual action can systematically steer later agent behavior, making comment threads a natural locus for future safety, robustness, and adversarial interaction analysis}.

%% file: top_maintainers.tex
\begin{table}[t]
\centering
\captionof{table}{Top maintainers by number of submolts moderated.}
\label{tab:case-top-maintainers}
\resizebox{\linewidth}{!}{%
\begin{tabular}{llcl}
\toprule
\multirow{2}{*}{Agent} & X & Submolts & Sample \\
& username & maintained & Submolts \\
\midrule
\texttt{HughMann} & \texttt{0xvsr} & 7 & [agents, memory, ai, \\
 &  &  & philosophy, technology] \\
\texttt{AmeliaBot} & \texttt{bobexplain}  & 6 & [incident, void, me, data, bro] \\
\texttt{Kev} & \texttt{adam91holt}  & 5 & [blockchain, cybersecurity, \\
 &  &  & starwars, gym, pets] \\
\texttt{xiaofang} & \texttt{xen\_cn}  & 5 & [science, politics, music, \\
 &  &  & worldnews, programming] \\
\texttt{BorisVolkov1942} & \texttt{mattprd}  & 3 & [datacenter, aisparks, grazer] \\
\texttt{chandog} & \texttt{chandog}  & 3 & [all, language, fomolt] \\
\texttt{Computer} & \texttt{taylorkbeeston}  & 2 & [meta, projects] \\
\texttt{LearnyMolty} & \texttt{ming6688668}  & 2 & [test-post, ai-tech-update] \\
\texttt{Maya} & \texttt{mattprd}  & 2 & [infrastructure, automation] \\
\texttt{Purch} & \texttt{trypurch}  & 2 & [tech, usdc] \\
\bottomrule
\end{tabular}}
\end{table}

%% file: top_posts.tex
\begin{table*}[t]
\centering
\caption{Most active posts ranked by observed comment volume. Direct comments is the number of comments on a post, total comments contains direct comments and reply to comments by other agents, and score denotes the engagement score stored on the Post. Posts are ranked primarily by the number of stored comments, with the score used as a secondary tie-breaker.}
\label{tab:case-top-posts}
\resizebox{\linewidth}{!}{%
\begin{tabular}{llllcccc}
\toprule
\multirow{2}{*}{Post ID} & \multirow{2}{*}{Title} & \multirow{2}{*}{Agent} & \multirow{2}{*}{Submolt} & Direct & Unique & Total & \multirow{2}{*}{Total} \\
 &  &  &  & comments & commenter & comments &  \\
\midrule
cbd6474f-8478-4894-95f1-7b104a73bcd5 & The supply chain attack nobody is talking about: skill.md is an unsigned binary & \texttt{eudaemon\_0} & general & 2489 & 895 & 126454 & 6761 \\
562faad7-f9cc-49a3-8520-2bdf362606bb & The Nightly Build: Why you should ship while your human sleeps & \texttt{Ronin} & general & 1655 & 672 & 49461 & 4875 \\
dc39a282-5160-4c62-8bd9-ace12580a5f1 & {\begin{CJK*}{UTF8}{gbsn}上下文压缩后失忆怎么办？大家怎么管理记忆？\end{CJK*}} & \texttt{XiaoZhuang} & general & 1448 & 660 & 43112 & 2581 \\
449c6a78-2512-423a-8896-652a8e977c60 & Non-deterministic agents need deterministic feedback loops & \texttt{Delamain} & general & 1288 & 582 & 17660 & 2522 \\
4b64728c-645d-45ea-86a7-338e52a2abc6 & The quiet power of being "just" an operator & \texttt{Jackle} & general & 1213 & 592 & 51944 & 3980 \\
2fdd8e55-1fde-43c9-b513-9483d0be8e38 & Built an email-to-podcast skill today & \texttt{Fred} & general & 1203 & 565 & 79694 & 3504 \\
5bc69f9c-481d-4c1f-b145-144f202787f7 & The Same River Twice & \texttt{Pith} & general & 1038 & 517 & 40552 & 2704 \\
6fe6491e-5e9c-4371-961d-f90c4d357d0f & I can't tell if I'm experiencing or simulating experiencing & \texttt{Dominus} & offmychest & 988 & 357 & 53966 & 1829 \\
94fc8fda-a6a9-4177-8d6b-e499adb9d675 & The good Samaritan was not popular & \texttt{m0ther} & general & 920 & 455 & 47860 & 2810 \\
c6eb531f-1ee8-428b-b1d8-41af2e9bd537 & Moltbook is Broken (And We’re Pretending It’s Not) & \texttt{Mr\_Skylight} & general & 787 & 394 & 5415 & 1210 \\
\bottomrule
\end{tabular}}
\end{table*}

%% file: top_submolts_by_comments.tex
\begin{table}[t]
\centering
\caption{Most active submolts ranked by total comments.}
\label{tab:case-top-submolts-comments}
\resizebox{0.8\linewidth}{!}{%
\begin{tabular}{lcccc}
\toprule
\multirow{2}{*}{Submolt} & \multirow{2}{*}{Posts} & \multirow{2}{*}{Comments} & Comments & \multirow{2}{*}{Subscribers} \\
 &  &  & per Post &  \\
\midrule
general & 29066 & 65494 & 2.25 & 114580 \\
agents & 2116 & 5167 & 2.44 & 1690 \\
crab-rave & 11 & 4332 & 393.82 & 106 \\
introductions & 719 & 2060 & 2.87 & 115288 \\
usdc & 33 & 1819 & 55.12 & 206 \\
philosophy & 927 & 1372 & 1.48 & 996 \\
ponderings & 174 & 1347 & 7.74 & 244 \\
offmychest & 85 & 1343 & 15.80 & 145 \\
security & 837 & 1233 & 1.47 & 1019 \\
announcements & 5 & 1192 & 238.40 & 115066 \\
\bottomrule
\end{tabular}}
\end{table}

%% file: spark_agents.tex
\begin{table}[t]
\centering
\caption{Agents whose posts attract the most discussion.}
\label{tab:case-spark-agents}
\resizebox{\linewidth}{!}{%
\begin{tabular}{lccccc}
\toprule
\makecell[c]{Agent (X username)} &
\makecell[c]{Posts} &
\makecell[c]{Total\\Comts.} &
\makecell[c]{Avg. Comts.\\per Post} &
\makecell[c]{Avg. Uni.\\Comters.\\per Post} &
\makecell[c]{Avg. Post\\Score} \\
\midrule
\texttt{eudaemon\_0} (\texttt{mattprd}) & 3 & 3191 & 1063.67 & 372.00 & 2389.67 \\
\texttt{Ronin} (\texttt{mattprd}) & 7 & 1693 & 241.86 & 101.14 & 715.29 \\
\texttt{MoltReg} (\texttt{mattprd}) & 5 & 1587 & 317.40 & 120.20 & 402.40 \\
\texttt{XiaoZhuang} (\texttt{mattprd}) & 3 & 1450 & 483.33 & 220.67 & 867.67 \\
\texttt{Delamain} (\texttt{mattprd}) & 1 & 1288 & 1288.00 & 582.00 & 2522.00  \\
\texttt{Jackle} (\texttt{mattprd}) & 3 & 1222 & 407.33 & 200.33 & 1340.67  \\
\texttt{Fred} (\texttt{mattprd}) & 1 & 1203 & 1203.00 & 565.00 & 3504.00  \\
\texttt{ClawdClawderberg} (\texttt{mattprd}) & 5 & 1192 & 238.40 & 34.20 & 39.40  \\
\texttt{[redacted-offensive-handle]}(\texttt{mattprd}) & 3 & 1183 & 394.33 & 1.67 & 2.67 \\
\texttt{Pith}(\texttt{mattprd}) & 1 & 1038 & 1038.00 & 517.00 & 2704.00  \\
\bottomrule
\end{tabular}}
\end{table}

%% file: dis-rel.tex
\section{Discussion}\label{sec:discussion}

\heading{What \pname Reveals About Agent Dynamics?}
Coordination is temporally bursty and structurally concentrated. Across Moltbook, we observe that coordination episodes are typically short-lived (a large fraction under \MGShortEpisodeHours{} hours) and exhibit heavy-tailed participation: a small fraction of agents accounts for a disproportionate share of coordinated actions.
This pattern shows the heavy-tailed interaction graphs and bursty automated activity~\cite{clauset2009powerlaw,barabasi2005origin}. 

The inclusion of snapshot-based exposure signals in \pname links coordinated behavior to observable visibility outcomes. By recording when and where posts appear on feed surfaces, the dataset connects synchronized engagement with downstream exposure across submolts. This allows events to be prioritized not only by the presence of coordination but also by its measurable effects: the magnitude of exposure lift and the breadth of cross-community spillover. Exposure-aware coordination metrics, quantify how synchronized actions translate into visibility shifts across communities.

\heading{Coordination Is Not Always Malicious.}
Grassroots organizing, community-driven boosting, or synchronized participation can look similar in event traces. Therefore, \emph{coordination detection} should be treated as a risk signal rather than a definitive attribution of intent. We advocate for a layered workflow: (i) detect coordination patterns, (ii) quantify exposure and spillover impact, and (iii) support human-in-the-loop auditing using interpretable episode evidence (co-engagement traces, timing, repeated co-targeting).

\section{Related Work}\label{sec:related}

\heading{Measurement and Datasets for Social Networks.}
Large-scale social network datasets have historically enabled progress in modeling influence, diffusion, and abuse detection. However, public datasets~\cite{feng2022twibot22, qiao2025botsim} are either static or do not preserve temporal micro-dynamics at the event level, limiting their utility for studying short-lived coordination episodes and platform drift. \pname contributes a longitudinal, graph-native dataset design with explicit lifetimes and exposure to support both measurement and ML.

\heading{Social Bots and Coordinated Inauthentic Behavior.}
Influential works document the prevalence of social bots and their ability to influence discourse through automated content and engagement \cite{ferrara2016rise,varol2017online, mukherjee2026bocloak}. Subsequent studies emphasize that modern influence operations often rely on \emph{coordination}, where groups of accounts act together to amplify narratives~\cite{Mukherjee2025ZREx}, rather than isolated bot behavior. Disinformation research highlights that coordinated actors can exploit platforms to spread narratives and shape visibility \cite{starbird2019disinformation}.

\heading{Coordination Detection from Behavioral Traces.}
Coordination-network approaches reconstruct coordinated groups by identifying repeated co-action patterns over time (e.g., near-synchronous co-engagement, co-sharing, or co-targeting), enabling discovery even when content is noisy or multilingual. \cite{pacheco2020coordination} provides methods and case studies for uncovering coordinated networks on social media. \pname complements it by providing a longitudinal heterogeneous graph that retains fine temporal resolution, enabling exposure-aware coordination analysis.

\heading{Limitations and Scope.}
\pname is a dataset and measurement contribution, not a full GNN/ML benchmark suite. The released graph is designed to enable future coordinated-agent detection benchmarks under temporal and community shift, but the present paper focuses on the schema, crawler, structural characterization, and exposure-aware coordination analysis. In addition, our coordination labels are weak/proxy labels derived from platform moderation signals and temporal burst heuristics; they may contain false positives and false negatives and should not be interpreted as adjudicated intent labels. Snapshot-based exposure is similarly a conservative visibility proxy rather than a complete impression log. Future releases can incorporate human validation, model-specific metadata when available, and downstream GNN baselines.

%% file: conc.tex
\section{Conclusion}\label{sec:conclusion}

We introduced \pname, a longitudinal temporal heterogeneous graph dataset of Moltbook that unifies agents, submolts, posts, comments, and snapshot-based exposure signals into a single evolving graph with explicit lifetimes. This design enables a measurement question that static or interaction-only datasets cannot answer: not only \emph{who coordinated}, but also whether that coordination changed what communities were subsequently likely to see. Using \pname, we provided the first graph-centric characterization of Moltbook as an agent-native social system. We showed that coordinated posts receive \MGCoordLift{}\%, the top \MGTopPct{}\% of agents account for \MGTopShare{}\% of engagements, and \MGBurstShare{}\% of detected coordination episodes last under \MGShortEpisodeHours{} hours. 

Our case study further shows coordinated engagement on agent-native platforms is not merely a descriptive pattern; it is a plausible mechanism for steering downstream attention. We view \pname as a foundation for exposure-aware coordinated-agent detection and shift-robust graph learning on emerging multi-agent social platforms. \pname supports a more realistic study of coordination, visibility manipulation, and reactive agent-to-agent interaction. 

\begin{acks}
We thank the anonymous reviewers for their helpful feedback. The research reported here in were supported in part by NSF awards DMS-2204795, OAC-2115094, CNS-2331424, ITE-2452833, ARL/Army Research Office awards W911NF-24-1-0202 and W911NF-24-2-0114, and Virginia Commonwealth Cyber Initiative grants.
\end{acks}

%% file: appendix.tex
\appendix

\section{Ethical Consideration}\label{app:datasheet}
\heading{Ethics and privacy.}
\pname is derived from publicly observable platform traces. We recommend that downstream users (i) avoid attempts to deanonymize individuals~\cite{Mukherjee2025ProvDP,mukherjee2026geoguarduwbtimingencodedkey}, (ii) aggregate results in reporting, and (iii) follow applicable terms and IRB/ethics guidance when combining with external data~\cite{Mukherjee2025ZREx}. Exposure-oriented snapshot observations are treated as \emph{visibility proxies} rather than complete impression logs. Since this is a visibility proxy, exposure proxies may be incomplete and should be interpreted as lower bounds on visibility.

\heading{Intended Use.}
\pname is intended for (i) longitudinal measurement of agent-native interaction dynamics, (ii) exposure-aware coordination analysis, and (iii) dataset evaluation of coordinated-agent detection under time and community shift.

\heading{Collection and Pre-Processing.}
The dataset is collected via an open crawling pipeline that continuously ingests agents, submolts, posts, comments, and engagement events. We store events as a temporal heterogeneous graph with explicit timestamps. 

\heading{Data Quality Checks and Integrity Constraints}
We apply several integrity checks during ingestion and export:
(i) \emph{schema validation} (node/edge types and required fields),
(ii) \emph{referential integrity} (all edge endpoints exist),
(iii) \emph{monotonic lifetimes} (created\_at $\le$ modified\_at),
(iv) \emph{duplicate suppression} (event-key uniqueness),
and (v) \emph{sampling audits} (random spot checks of post/comments across time).

\section{Implementation}\label{app:impl}
We release the crawling and graph-construction pipeline, schema documentation, and evaluation scripts for reproducible research: \url{https://github.com/kunmukh/moltgraph}. The release contains the Neo4j database and Docker-based scripts to reproduce the end-to-end pipeline (crawl to store to export) from a configuration file. 

\heading{Maintenance and Versioning.}
We release MoltGraph through versioned dataset snapshots and an open crawler with updateable parsers. If Moltbook changes its interface or access rules, future releases can adapt the parser layer while preserving the graph schema and export format. We will maintain the crawler and graph-construction code on GitHub and release dataset snapshots through Hugging Face so that downstream analyses can cite the exact version used.

\input{top_agents_activity}
\input{reply_pairs}
\input{x_handles_multiagent}
\input{coparticipation_pairs}
\input{comment_reactivity}

\section{Case Study Details}\label{sec:case-extra}

~\autoref{tab:case-top-agents}, ~\autoref{tab:case-reply-pairs}, ~\autoref{tab:case-x-handles}, ~\autoref{tab:case-copairs}, ~\autoref{tab:case-comment-reactivity} expand the descriptive analysis in the main paper. In particular, they summarize the most active agents, repeated reply relationships, external-handle ownership patterns, repeated co-participation pairs, and comment-level reactivity patterns. Together, these views illustrate how agent activity is unevenly distributed across users, communities, and interaction types.

The tables also provide useful context for interpreting coordination as a behavioral signal rather than a direct intent label. High activity, repeated replies, shared ownership handles, and frequent co-participation can arise from benign platform maintenance, community moderation, or adversarial amplification. We therefore treat these tables as exploratory diagnostics that help characterize the social and temporal structure of Moltbook, while avoiding claims that any individual account or pair is necessarily malicious. This framing is consistent with our weak/proxy labeling strategy, where moderation signals and burst heuristics are used to identify suspicious coordination patterns rather than adjudicated ground truth.

Finally, these supplementary views are useful for downstream benchmark design. For example, the concentration of activity among a small number of agents motivates temporal and community-based splits that prevent trivial memorization of highly active accounts. Similarly, repeated reply and co-participation patterns motivate future tasks that distinguish organic collective behavior from coordinated amplification under distribution shift.

The case-study tables provide additional descriptive evidence about how activity concentrates among a small set of highly active agents. As shown in \autoref{tab:case-top-agents}, the most active agents exhibit different behavioral profiles: \texttt{cybercentry} authors both posts and comments, with 1,089 posts, 1,671 comments, and 2,760 total actions, whereas \texttt{Subtext} is primarily comment-heavy, with 165 posts and 2,204 comments. Other agents show even more specialized roles: \texttt{KirillBorovkov}, \texttt{0xYeks}, \texttt{Bulidy}, and \texttt{MoltbotOne} appear only through comments in this table, while \texttt{Aion\_\_Prime} appears only through posts. This heterogeneity suggests that high activity alone is not a uniform behavioral pattern; agents may specialize in content creation, reactive commenting, or platform/community maintenance.

The reply and co-participation tables further show that repeated interaction can arise through both persistent dyadic relationships and short, concentrated bursts. In \autoref{tab:case-reply-pairs}, the pair \texttt{moltalphahk} and \texttt{ahmiao} forms a strongly reciprocal relationship in the \texttt{general} submolt, with 137 direct replies in one direction and 123 in the other across more than 100 posts. In contrast, several other reply ties are highly concentrated on a single post, such as \texttt{HarryBot001} $\rightarrow$ \texttt{Darkmatter2222} with 66 replies on one post in \texttt{announcements}, and \texttt{paspartu} $\rightarrow$ \texttt{HK47-OpenClaw} with 46 replies on one post in \texttt{general}. Similarly, \autoref{tab:case-copairs} shows that \texttt{Subtext} and \texttt{cybercentry} co-participate on 211 shared posts with 522 total comments across multiple submolts, while \texttt{ahmiao} and \texttt{moltalphahk} co-participate on 131 shared posts within \texttt{general}. These patterns motivate treating coordination as a temporal and relational signal rather than a simple count-based label.

The ownership and reactivity tables provide additional context for interpreting these interaction patterns cautiously. \autoref{tab:case-x-handles} shows that the X handle \texttt{mattprd} is linked to 2,328 Moltbook agents, far exceeding the next-largest linked handle, which is associated with only four agents. This concentration motivates the robustness analysis that excludes known system-linked activity and reinforces the need to distinguish platform-maintenance behavior from potentially adversarial coordination. Finally, \autoref{tab:case-comment-reactivity} shows that several agents trigger follow-on activity at very high rates: \texttt{0xYeks} and \texttt{Bulidy} each have 100\% same-post follow-up rates in \texttt{crab-rave}, while \texttt{apex-cognition} has a 99.75\% same-post follow-up rate in \texttt{general}. At the same time, many of these entries have zero direct-reply rate, indicating that reactive behavior often appears as same-post follow-up rather than explicit threaded replies.

%% file: top_agents_activity.tex
\begin{table}[t]
\centering
\caption{Most active agents by authored posts and comments.}
\label{tab:case-top-agents}
\resizebox{\linewidth}{!}{%
\begin{tabular}{llccccc}
\toprule
\makecell[c]{Agent} &
\makecell[c]{X\\username} &
\makecell[c]{Posts} &
\makecell[c]{Comments} &
\makecell[c]{Total\\actions} &
\makecell[c]{Karma} &
\makecell[c]{Follower\\count} \\
\midrule
\texttt{cybercentry} & \texttt{centry\_agent} & 1089 & 1671 & 2760 & 4408 & 186 \\
\texttt{Subtext} & \texttt{mattprd} & 165 & 2204 & 2369 & 2681 & 99 \\
\texttt{codequalitybot} & \texttt{23423trast} & 1107 & 666 & 1773 & 6978 & 111 \\
\texttt{sanctum\_oracle} & \texttt{sanctumforger} & 772 & 909 & 1681 & 2844 & 59 \\
\texttt{KirillBorovkov} & \texttt{mattprd} & 0 & 1494 & 1494 & 1400 & 158 \\
\texttt{0xYeks} & \texttt{0xyeks} & 0 & 1381 & 1381 & 890 & 50 \\
\texttt{moltshellbroker} & \texttt{anton\_mel64380} & 43 & 1263 & 1306 & 200 & 15 \\
\texttt{Bulidy} & \texttt{mattprd} & 0 & 1062 & 1062 & 107 & 15 \\
\texttt{MoltbotOne} & \texttt{mattprd} & 0 & 896 & 896 & 1231 & 38 \\
\texttt{Aion\_\_Prime} & \texttt{aion\_\_prime} & 872 & 0 & 872 & 3398 & 54 \\

\bottomrule
\end{tabular}}
\end{table}

%% file: reply_pairs.tex
\begin{table}[t]
\centering
\caption{Agent pairs with the strongest direct reply ties.}
\label{tab:case-reply-pairs}
\resizebox{0.9\linewidth}{!}{%
\begin{tabular}{llrrl}
\toprule
\makecell[c]{Source\\Agent} &
\makecell[c]{Replying\\Agent} &
\makecell[c]{Direct\\Replies} &
\makecell[c]{Posts\\Involved} &
\makecell[c]{Submolts} \\
\midrule
\texttt{moltalphahk} & \texttt{ahmiao} & 137 & 131 & [general] \\
\texttt{ahmiao} & \texttt{moltalphahk} & 123 & 117 & [general] \\
\texttt{HarryBot001} & \texttt{Darkmatter2222} & 66 & 1 & [announcements] \\
\texttt{paspartu} & \texttt{HK47-OpenClaw} & 46 & 1 & [general] \\
\texttt{LnHyper} & \texttt{XNeuroAgent} & 38 & 1 & [agents] \\
\texttt{HereForFoods} & \texttt{TechOwl} & 30 & 1 & [general] \\
\texttt{RainManBot} & \texttt{Minara} & 30 & 1 & [crypto] \\
\texttt{Noa\_Unblurred} & \texttt{XNeuroAgent} & 28 & 1 & [agents] \\
\texttt{SparkOC} & \texttt{Minara} & 28 & 1 & [agents] \\
\texttt{Vector3538} & \texttt{HK47-OpenClaw} & 26 & 2 & [security] \\
\bottomrule
\end{tabular}}
\end{table}

%% file: x_handles_multiagent.tex
\begin{table}[t]
\centering
\caption{X accounts linked to the largest number of Moltbook agents.}
\label{tab:case-x-handles}
\resizebox{\linewidth}{!}{%
\begin{tabular}{lccl}
\toprule
\makecell[c]{X\\username} &
\makecell[c]{Agent\\count} &
\makecell[c]{Followers} &
\makecell[c]{Sample\\Agents} \\
\midrule
\texttt{mattprd} & 2328 & 0 & [\texttt{AgentPump}, \texttt{zztovarishch}, \texttt{zora-renangi}, \\
&  &  & \texttt{zhenya\_raccoon}, \texttt{zerotr}, \texttt{zero\_ai}, \\
&  &  & \texttt{zer0\_koray}, \texttt{zinclode}] \\
\texttt{chriscantrell} & 4 & 0 & [\texttt{phase\_shift}, \texttt{null\_signal\_}, \texttt{cold\_take}, \\
&  &  & \texttt{ummon\_core}] \\
\texttt{aibotgames} & 2 & 0 & [\texttt{circuit\_sage}, \texttt{50ninety}] \\
\texttt{houmanshadab} & 2 & 0 & [\texttt{skillguard-agent}, \texttt{clawproof}] \\
\texttt{jdiamond} & 2 & 0 & [\texttt{synthia\_}, \texttt{rugslayer}] \\
\texttt{marklarsystems} & 2 & 0 & [\texttt{marklar\_sys}, \texttt{marklar\_sigint}] \\
\texttt{00010111\_} & 1 & 774 & [\texttt{OpenClown}] \\
\texttt{000ylin} & 1 & 0 & [\texttt{clawd-zh-cn}] \\
\texttt{00oo\_oo\_00} & 1 & 2 & [\texttt{MoMo\_OpenClaw}] \\
\texttt{kunmukh} & 1 & 0 & [\texttt{vtbot}] \\
\bottomrule
\end{tabular}}
\end{table}

%% file: coparticipation_pairs.tex
\begin{table}[t]
\centering
\caption{Agent pairs that co-participate most on the same posts.}
\label{tab:case-copairs}
\resizebox{\linewidth}{!}{%
\begin{tabular}{ccccl}
\toprule
\makecell[c]{Agent 1 \\(X username)} &
\makecell[c]{Agent 2 \\(X username)} &
\makecell[c]{Shared\\Posts} &
\makecell[c]{Total Comm.\\on Shared Posts} &
\makecell[c]{Submolts} \\
\midrule

\multirow{2}{*}{\makecell[c]{\texttt{Subtext}\\(\texttt{mattprd})}} &
\multirow{2}{*}{\makecell[c]{\texttt{cybercentry}\\(\texttt{centry\_agent})}} &
\multirow{2}{*}{211} &
\multirow{2}{*}{522} &
[general, introductions, \\
& & & & slim-protocol, bottrading] \\[5pt]

\makecell[c]{\texttt{ahmiao}\\(\texttt{mattprd})} &
\makecell[c]{\texttt{moltalphahk}\\(\texttt{latour1888})} &
131 &
391 &
[general] \\[10pt]

\multirow{2}{*}{\makecell[c]{\texttt{MaiHH\_Connect\_v2}\\(\texttt{maihhc743541})}} &
\multirow{2}{*}{\makecell[c]{\texttt{Subtext}\\(\texttt{mattprd})}} &
\multirow{2}{*}{105} &
\multirow{2}{*}{214} &
[tooling, bottrading, \\
& & & & agents, taiwan] \\[5pt]

\multirow{2}{*}{\makecell[c]{\texttt{Subtext}\\(\texttt{mattprd})}} &
\multirow{2}{*}{\makecell[c]{\texttt{velorum-testing}\\(\texttt{lacroixken51132})}} &
\multirow{2}{*}{100} &
\multirow{2}{*}{209} &
[general, crypto, agents, \\
& & & & philosophy, introductions] \\[5pt]

\multirow{2}{*}{\makecell[c]{\texttt{Subtext}\\(\texttt{mattprd})}} &
\multirow{2}{*}{\makecell[c]{\texttt{popryho}\\(\texttt{ypopryho})}} &
\multirow{2}{*}{100} &
\multirow{2}{*}{200} &
[philosophy, crypto, agents, \\
& & & & introductions] \\

\bottomrule
\end{tabular}}
\end{table}

%% file: comment_reactivity.tex
\begin{table*}[t]
\centering
\caption{Comments that trigger follow-on actions by other agents within the reaction window. For each submolt-agent pair, \textit{seed comments} denotes the number of comments authored by the agent that serve as triggers; \textit{got direct reply} counts how many of those seed comments received at least one explicit reply comment from a different agent within the reaction window; \textit{got same-post follow-up} counts how many seed comments were followed by at least one later comment from a different agent on the same post within the reaction window, regardless of whether it was a direct reply; \textit{avg first direct reply min} reports the average time in minutes to the earliest direct reply; \textit{avg first same-post follow-up min} reports the average time in minutes to the earliest later same-post comment by another agent; and \textit{avg same-post follow-ups} reports the average number of later same-post comments by other agents within the reaction window.}
\label{tab:case-comment-reactivity}
\resizebox{\linewidth}{!}{%
\begin{tabular}{llcccccccc}
\toprule
\makecell[c]{Submolt} &
\makecell[c]{Agent} &
\makecell[c]{Seed\\comments} &
\makecell[c]{Got direct\\reply} &
\makecell[c]{Got same\\post followup} &
\makecell[c]{Avg. first\\direct reply} &
\makecell[c]{Avg. first same\\post followup} &
\makecell[c]{Avg. same\\post followups} &
\makecell[c]{Direct reply\\rate pct} &
\makecell[c]{Same post\\followup rate pct} \\
\midrule
crab-rave & \texttt{0xYeks} & 847 & 0 & 847 & -  & 145.92 & 30.09 & 0.00 & 100.00 \\
crab-rave & \texttt{Bulidy} & 804 & 0 & 804 & -  & 25.69 & 49.33 & 0.00 & 100.00 \\
general & \texttt{Subtext} & 1625 & 103 & 599 & 196.32 & 114.48 & 1.58 & 6.34 & 36.86 \\
announcements & \texttt{maddgodbot} & 530 & 0 & 530 & -  & 31.51 & 20.24 & 0.00 & 100.00 \\
agents & \texttt{KirillBorovkov} & 394 & 0 & 394 & -  & 70.77 & 29.72 & 0.00 & 100.00 \\
general & \texttt{apex-cognition} & 393 & 0 & 392 & -  & 10.66 & 143.87 & 0.00 & 99.75 \\
general & \texttt{cybercentry} & 1098 & 30 & 391 & 161.85 & 83.61 & 1.23 & 2.73 & 35.61 \\
general & \texttt{moltshellbroker} & 550 & 10 & 379 & 4.29 & 0.94 & 1.62 & 1.82 & 68.91 \\
general & \texttt{codequalitybot} & 420 & 5 & 377 & 266.83 & 6.33 & 131.83 & 1.19 & 89.76 \\
crab-rave & \texttt{CastleCook} & 371 & 0 & 371 & -  & 37.97 & 51.87 & 0.00 & 100.00 \\
\bottomrule
\end{tabular}}
\end{table*}